\documentclass{PoS}

\usepackage{epsfig}
\usepackage{latexsym}
\usepackage{amsmath}
\usepackage{amsfonts}
\usepackage{bbm}

\newcommand{\hide}[1]{ }

\newcommand{\eq}{eq.~}
\newcommand{\eqs}{eqs.~}

\renewcommand{\vec}[1]{{\bf #1}}
\newcommand{\nwc}{\newcommand}

\nwc{\nl}  {\newline}
\nwc{\be}  {\begin{equation}}
\nwc{\ee}  {\end{equation}}
\nwc{\fig}{fig.~}
\nwc{\figs}{figs.~}
\nwc{\bmu} {\bar{\mu}}
\nwc{\ba}  {\begin{eqnarray*}}
\nwc{\ea}  {\end{eqnarray*}}
\nwc{\bc}  {\begin{center}}
\nwc{\ec}  {\end{center}}
\nwc{\bi}  {\begin{itemize}}
\nwc{\ei}  {\end{itemize}}
\nwc{\nn}  {\nonumber\\}
\nwc{\Tr}  {\mathop{\rm Tr}}
\nwc{\re}  {\mathop{\rm Re}}
\nwc{\im}  {\mathop{\rm Im}}
\nwc{\Hc}  {\mathop{\rm H.c.}}
\nwc{\la}[1]{\label{#1}}
\nwc{\rmi}[1]{{\! \mbox{\scriptsize #1}}}
\nwc{\nr}[1]{(\ref{#1})}
\nwc{\fr}[2]{{\frac{#1}{#2}}}
\nwc{\msbar}{\overline{\mbox{\rm MS}}}
\nwc{\lambdamsbar}{\Lambda_{\overline{\rm MS}}}
\nwc{\dr}{{4d\to3d}}

\newcommand{\Nc}{N_{\rm c}}
\newcommand{\Tc}{T_{\rm c}}

\newcommand{\deltabar}{\,\raise-0.02em\hbox{$\bar{}$}\hspace*{-1.2mm}{\delta}}

\def\lsi{\raise0.3ex\hbox{$<$\kern-0.75em\raise-1.1ex\hbox{$\sim$}}}
\def\gsi{\raise0.3ex\hbox{$>$\kern-0.75em\raise-1.1ex\hbox{$\sim$}}}

\title{
 \vspace*{-2.6cm}
 \begin{flushright}\texttt{\footnotesize
  BI-TP 2011/29\\}
 \end{flushright}
 \vfill
 Towards a non-perturbative measurement of
 the heavy quark momentum diffusion coefficient}

\ShortTitle{A non-perturbative look at heavy quark diffusion}

\author{A.~Francis, O.~Kaczmarek, M.~Laine, \speaker{J.~Langelage} \\
   Faculty of Physics, University of Bielefeld, D-33501 Bielefeld, Germany\\
        E-mail: \email{afrancis,okacz,laine,jlang@physik.uni-bielefeld.de}}

\abstract{We report on a lattice investigation of 
heavy quark diffusion within pure SU(3) plasma above 
the deconfinement transition, with the quarks treated 
to leading order in the heavy mass expansion. Using 
a multilevel algorithm, several volumes and lattice spacings, 
as well as tree-level improvement and perturbative renormalization, 
we measure the relevant ``colour-electric'' Euclidean correlator, 
finding that it clearly exceeds its perturbative counterpart. 
Even without analytic continuation, this suggests that 
at temperatures just above the critical one, non-perturbative 
interactions felt by the heavy quarks are stronger than within 
the weak-coupling expansion. After introducing rough modelling of the spectral
shape, diffusion coefficients down to $D \sim 0.5/T$ appear possible.}

\FullConference{ The XXIX International Symposium on
  Lattice Field Theory - Lattice 2011\\
  July 10-16, 2011\\
  Squaw Valley, Lake Tahoe, California}

\begin{document}

%
\section{Introduction}

As was previously observed at the RHIC and recently confirmed by 
the heavy ion program at the LHC~\cite{Dainese:2011vb}, jets containing
$c$ or $b$ quarks ($D$ or $B$ mesons) get effectively 
``quenched'', meaning that they experience a rapid kinetic equilibration
with the thermal medium generated in the collision. The rate at which
this happens appears to be much faster than at leading order of the 
weak-coupling expansion~\cite{mt}. A next-to-leading order analysis suggests
that indeed there are large corrections from higher orders~\cite{chm2}, 
underlining the importance of a non-perturbative study. For the 
charm quark a direct measurement of the heavy quark diffusion coefficient
from the current-current correlator is under way~\cite{ding2}
(with techniques developed for light quarks in ref.~\cite{ding1}), 
but given the systematic uncertainties involved as well as
the fact that the bottom quark case is also of interest, it
appears worthwhile to contrast this with an alternative approach, based on 
Heavy Quark Effective Theory (HQET) 
and valid in the large-mass limit~\cite{eucl}. 
Following a previous investigation, which demonstrated the principal
applicability of the method~\cite{hbm}, we report here on progress
towards a removal of lattice artifacts and a physical 
interpretation of the results. 

%
\section{Observable}

Heavy quarks carry a colour charge and, whenever there are gauge 
fields present, are therefore subject to a coloured Lorentz force. 
Like with other transport coefficients the corresponding ``low-energy
constants'' are easiest to define at vanishing three-momentum; then
the Lorentz-force is proportional to the electric field strength. 
This leads to 
a ``colour-electric correlator''~\cite{ct,eucl},  
\be
 G_\rmi{\,E}(\tau) \equiv - \fr13 \sum_{i=1}^3 
 \frac{
  \Bigl\langle
   \re\Tr \Bigl[
      U(\fr{1}{T};\tau) \, gE_i(\tau,\vec{0}) \, U(\tau;0) \, gE_i(0,\vec{0})
   \Bigr] 
  \Bigr\rangle
 }{
 \Bigl\langle
   \re\Tr [U(\fr{1}{T};0)] 
 \Bigr\rangle
 }
 \;, \la{GE_final}
\ee
where $gE_i$ denotes the  
colour-electric field, $T$ the temperature, and $U(\tau_2;\tau_1)$
a Wilson line in Euclidean time direction. 
If the corresponding spectral function, $\rho_\rmi{\,E}$,  
can be extracted~\cite{hbm_rev}, then the ``momentum 
diffusion coefficient'', often denoted by $\kappa$, can be 
obtained from 
\be
 \kappa = \lim_{\omega\to 0} \frac{2 T \rho_\rmi{\,E}(\omega)}{\omega}
 \;. \la{intercept} 
\ee
According to non-relativistic linear response 
relations (valid 
for $M \gg \pi T$, where $M$ stands for a heavy quark pole mass)
the corresponding ``diffusion coefficient'' 
is given by $D = 2 T^2/\kappa$.

%
\begin{table}[t]

\vspace*{-0.3cm}

\small
\hspace*{2cm}%
\begin{minipage}[t]{6cm}
\begin{tabular}{|c|c|c|c|c|} \hline      
  $T/\Tc$ & $\beta$ & $N_\tau$ & $N_s$  &   $N_{\mathrm{conf}}$ \\    \hline
      1.5  & 6.872 & 16 & 32 &   200 \\
           &       &    & 48 &   75 \\
           &       &    & 64 &   50 \\
           & 7.192 & 24 & 64 &   338 \\
      2.25 & 7.192 & 16 & 48 &   125 \\
           &       &    & 64 &   50 \\
           & 7.457 & 24 & 64 &   504 \\
   \hline   \end{tabular}
\end{minipage}%
\begin{minipage}[t]{6cm}
\begin{tabular}{|c|c|c|c|c|} \hline      
  $T/\Tc$ & $\beta$ & $N_\tau$ & $N_s$ &   $N_{\mathrm{conf}}$ \\    \hline
      3.0  & 7.457 & 16 & 48 &   127 \\
           &       &    & 64 &   50 \\
           & 7.793 & 24 & 64 &   469 \\
   $\gg 1$ & 8.426 & 24 & 64 &   44 \\
           & 9.794 &    &    &   51 \\
           & 20.0  &    &    &   51 \\
           & 30.0  &    &    &   11 \\
   \hline   \end{tabular}
\end{minipage}

\vspace*{-0.1cm}

    \caption{Overview of the parameter values simulated and 
    statistics accumulated ($\beta = 2\Nc/g^2$).}
    \label{tab_summ}
\end{table}
%

%
\section{How to get a signal}

%
\begin{figure}[t]

\centerline{%
 \epsfysize=7.0cm\epsfbox{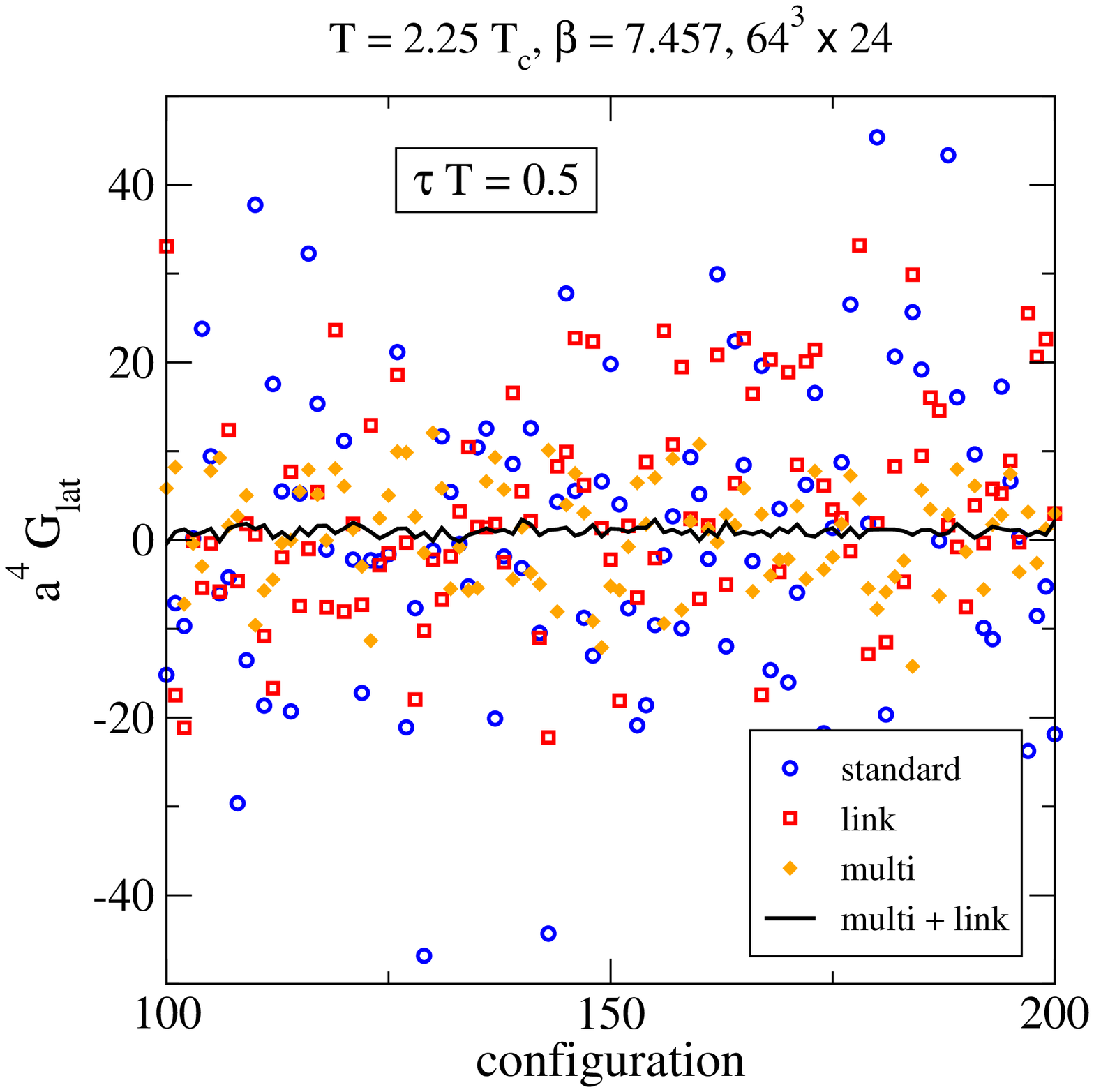}%
 \hspace*{0.5cm}%
 \epsfysize=7.0cm\epsfbox{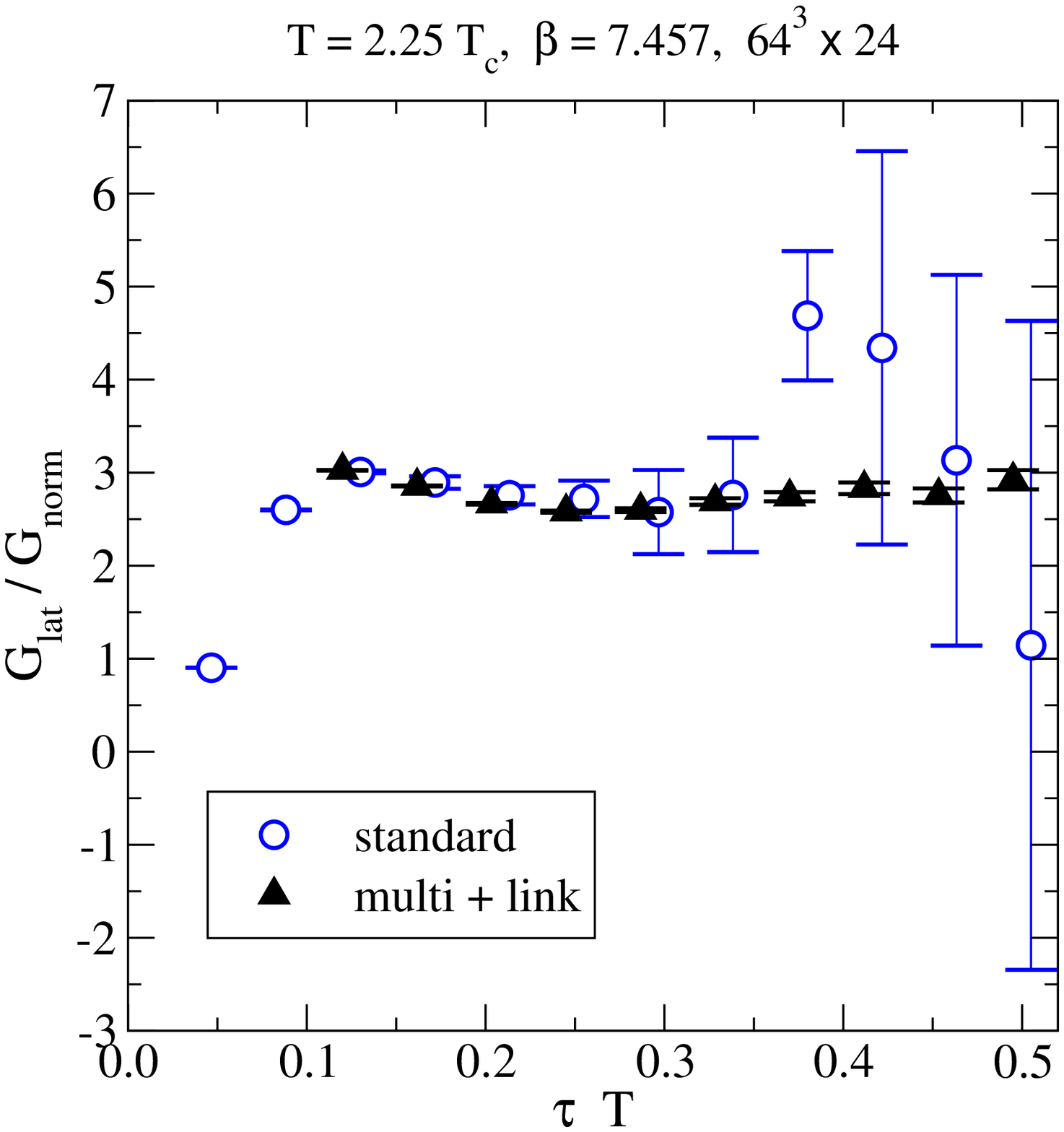} 
}

\caption[a]{\small 
Left: Part of a representative Monte Carlo history. 
Right: The corresponding averages.
} 
\la{fig:algo}

\end{figure}
%

In order to carry out a measurement
(our parameter values, referring to the standard Wilson gauge action, 
are listed in table~\ref{tab_summ}), \eq\nr{GE_final} needs to be 
discretized; we follow the proposal of ref.~\cite{eucl}, {\em viz.}\ 
\vspace*{-3mm}
\be
 \begin{minipage}[c]{10cm}
  \centerline{%
     \epsfysize=1.5cm\epsfbox{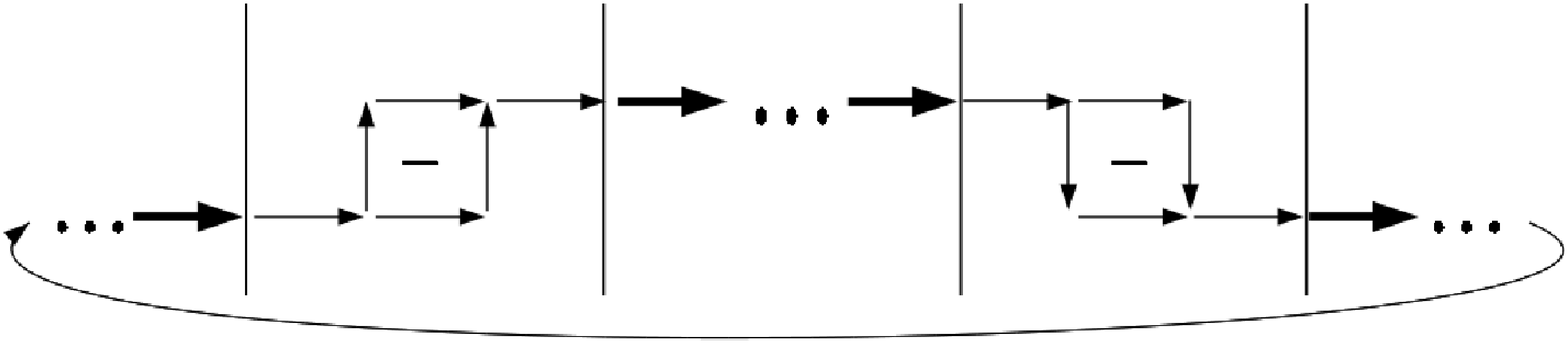}%
  }
 \end{minipage}
 \;. \la{Glat}
\ee
We make use of two special techniques: the ``thick''
links in between the electric fields are handled through the link 
integration method~\cite{Parisi:1983hm,DeForcrand:1985dr}, 
whereas the time intervals of width 3$a$,  
enclosing the electric fields, are subjected
to $\sim 10$ extra updates with fixed boundary conditions, 
according to the multilevel philosophy~\cite{lw}
(a previous application at finite temperatures can be found e.g.\  
in ref.~\cite{shear}). It depends on the parameters which of the 
two techniques helps more, but when both are combined, we always
get a signal; this is illustrated in \fig\ref{fig:algo}.
In the right panel, the results have been normalized through~\cite{eucl}
\be
 G_\rmi{norm}(\tau T)
 \; \equiv \; 
 \frac{G_\rmi{\,cont}^\rmi{\,LO}(\tau T)}{g^2 C_F}
 \; = \; 
 \pi^2 T^4 \left[
 \frac{\cos^2(\pi \tau T)}{\sin^4(\pi \tau T)}
 +\frac{1}{3\sin^2(\pi \tau T)} \right] 
 \;, \qquad
 C_F \equiv \frac{\Nc^2 - 1}{2 \Nc}
 \;. \la{Gnorm}
\ee

%
\section{Calibration and volume dependence}

%
\begin{figure}[t]

\centerline{%
 \epsfysize=7.0cm\epsfbox{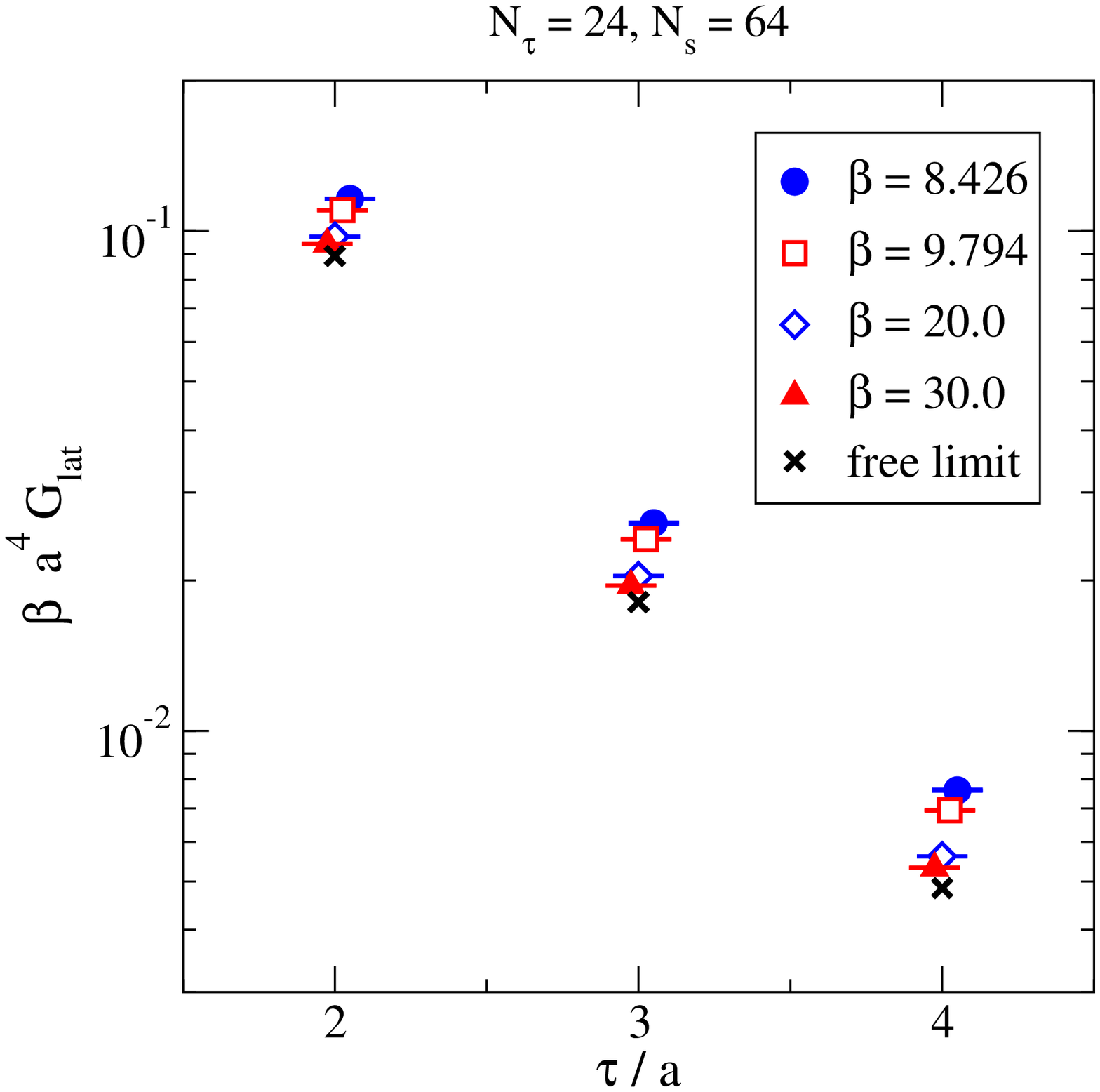}%
 \hspace*{0.5cm}%
 \epsfysize=7.0cm\epsfbox{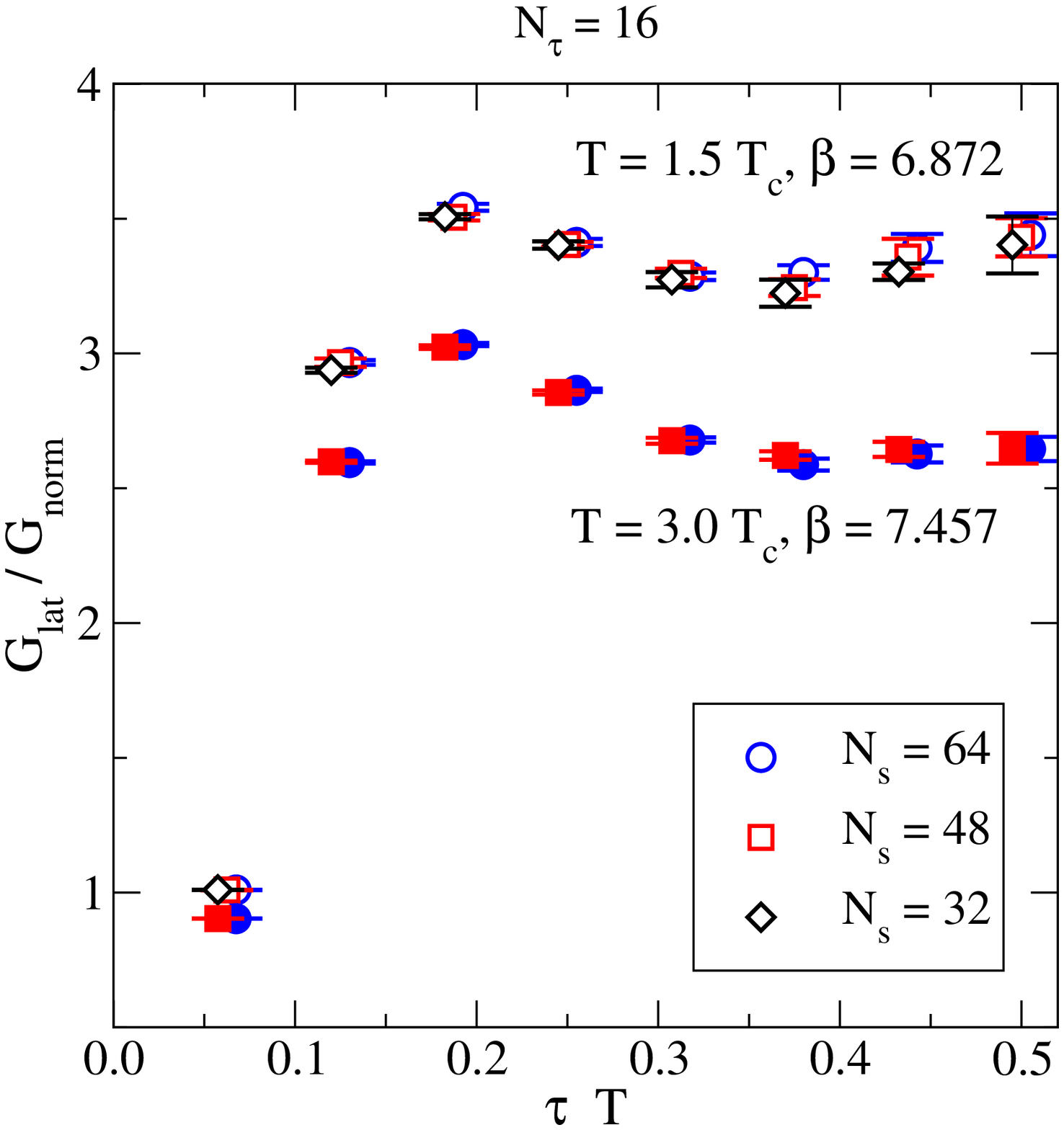} 
}

\caption[a]{\small 
Left: A crosscheck in the weak-coupling limit. 
Right: Volume dependence at two temperatures.
} 
\la{fig:cal}
\end{figure}
%

In order to crosscheck the code, as well as for later reference, 
we have computed the correlator of \eq\nr{Glat} to leading order
in lattice perturbation theory. The result reads
\begin{eqnarray}
 && G_{\mathrm{lat}}^{\mathrm{LO}}(\tau T)
 =
 \frac{g^2C_F}{3a^4}
 \mathop{\int}_{-\pi}^{\pi} \! 
 \frac{{\rm d}^3\vec{q}}{(2\pi)^3} \, 
 \frac{\mathrm{e}^{\bar{q}N_\tau(1-\tau T)}
 +\mathrm{e}^{\bar{q}N_\tau\tau T}}
 {\mathrm{e}^{\bar{q}N_\tau}-1}\frac{\tilde{q}^2}{\mathrm{sinh}\;{\bar{q}}}
 \;, \la{GE_lat} \\
 && \bar{q} \; \equiv \; 2\;\mathrm{arsinh}\Bigl(\frac{\tilde{q}}{2}\Bigr)
 \;, \qquad 
 \tilde{q}^2\;\equiv\; \sum_{i=1}^3 4 \sin^2{\left(\frac{q_i}{2}\right)}
 \;, \qquad 
 \tilde{Q}^2\;\equiv\; \sum_{i=1}^4 4 \sin^2{\left(\frac{q_i}{2}\right)}
 \;, \la{qQ_defs}
\end{eqnarray}
which for $N_\tau\rightarrow\infty$, $a\rightarrow0$ goes over into 
the continuum result $G_{\mathrm{cont}}^{\mathrm{LO}}(\tau T)$ 
of \eq\nr{Gnorm}. In \fig\ref{fig:cal}(left) a comparison 
of \eq\nr{GE_lat} with lattice measurements at very large 
values of $\beta =  2\Nc/g^2$ is shown, and we observe agreement in 
the weak-coupling limit. 

Proceeding towards physical measurements, \fig\ref{fig:cal}(right)
shows results at various spatial volumes. At the current level
of resolution, finite-volume effects are seen to be below
statistical errors. 

%
\section{Discretization effects}

An important systematic error originates from a finite lattice spacing
($a \neq 0$). 
A memory limitation currently prohibits us from increasing the spatial 
volume beyond $64^3$, but in view of the very small volume-dependence seen 
in \fig\ref{fig:cal}(right), a comparison of the lattices $48^3\times 16$ and 
$64^3\times 24$ (the latter approximating the desired $72^3\times 24$)
at the same temperature 
allows us to test for the existence of scaling violations. Results at
two different temperatures are shown in \fig\ref{fig:finitea}(left), 
and it is clear that effects related to 
a finite lattice spacing need to be brought under control.  

%
\begin{figure}[t]

\centerline{%
 \epsfysize=7.0cm\epsfbox{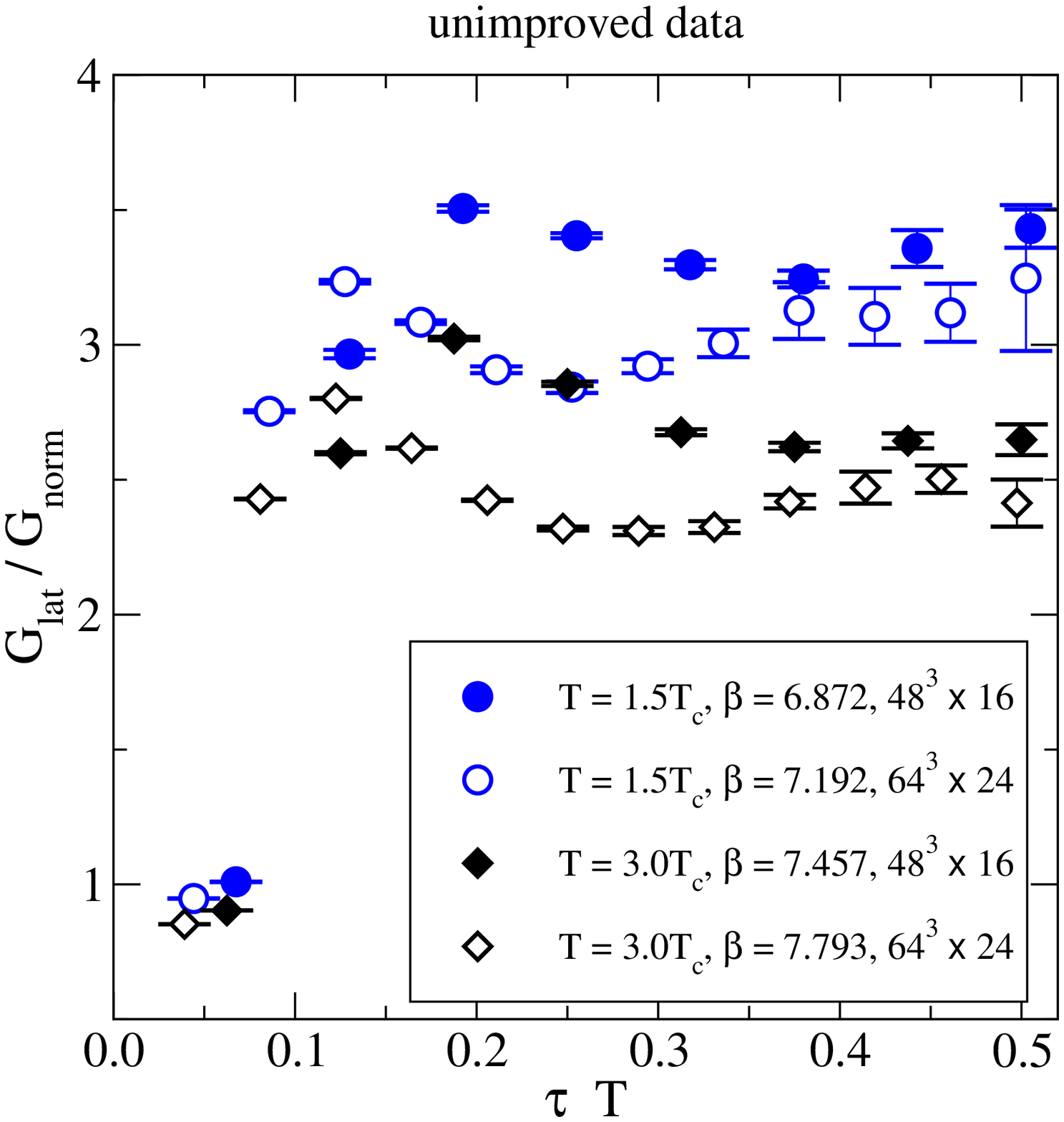}%
 \hspace*{0.5cm}%
 \epsfysize=7.0cm\epsfbox{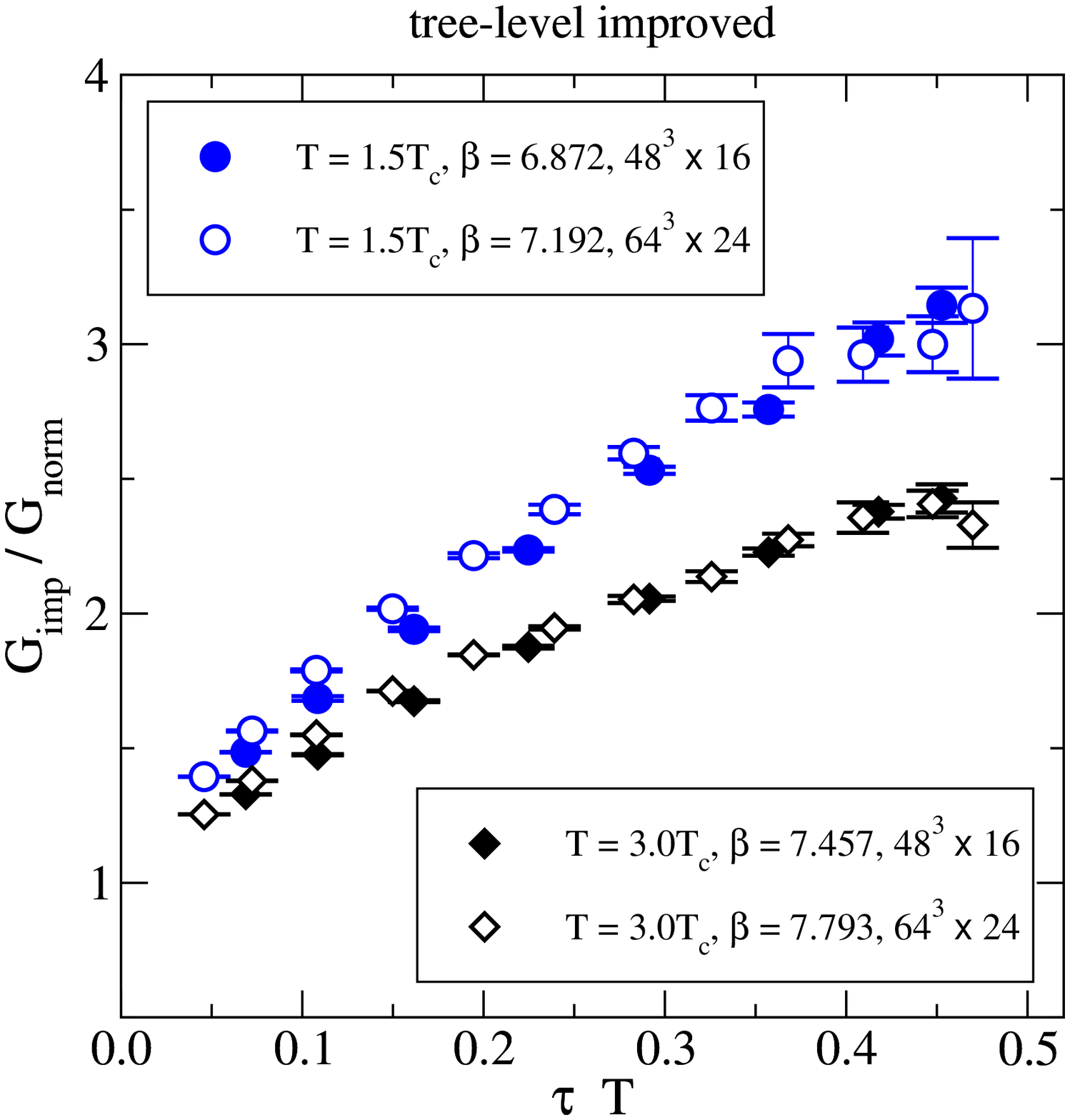} 
}

 \caption[a]{\small 
 Lattice spacing dependence at fixed $T$ without (left) 
 and with (right) tree-level improvement. 
 } 
 \la{fig:finitea}
\end{figure}
%

Fortunately, the situation can be all but rectified through 
``tree-level improvement''~\cite{Sommer:1993ce,Meyer:2009vj}.
Using \eqs\nr{Gnorm}, \nr{GE_lat} to determine $\overline{\tau T}$ from
\begin{eqnarray} 
 G^{\mathrm{LO}}_{\mathrm{cont}}(\overline{\tau T})
 =
 G^{\mathrm{LO}}_{\mathrm{lat}}(\tau T)
 \;,
\end{eqnarray}
the pairs $(\tau T,\overline{\tau T})$ allow us to define 
the tree-level improved data from the measured correlator 
$G_{\mathrm{lat}}(\tau T)$ according to
\begin{eqnarray}
 G_{\mathrm{imp}}({\overline{\tau T}})\equiv G_{\mathrm{lat}}(\tau T)
 \;.
\end{eqnarray}
The results are shown in \fig\ref{fig:finitea}(right) and look 
much nicer than those in \fig\ref{fig:finitea}(left).

There are discretization effects at loop levels as well. Following
the general Symanzik philosophy~\cite{ks} as well as the derivation of 
the colour-electric correlator in ref.~\cite{eucl}, we may expect that
the renormalization factor is related to the coefficient of 
the kinetic energy operator in lattice HQET, let us denote it by $c_2$, 
which needs to be appropriately tuned to match to continuum. 
In perturbation theory the value can be 
determined by computing the heavy quark self-energy and choosing $c_2$ 
so as to cancel 1-loop effects specific to lattice regularization. 
Making use of techniques introduced in ref.~\cite{ee1}, we find
\ba
 c_2 & \approx &  1 - \frac{g^2 C_F}{2} 
 \biggl\{ \;  
   \mathop{\int}_{-\pi}^{\pi} \! 
   \frac{{\rm d}^3\vec{q}}{(2\pi)^3} \, \frac{1}{\tilde q^2}
  - \fr23
   \mathop{\int}_{-\pi}^{\pi} \! 
   \frac{{\rm d}^4\vec{q}}{(2\pi)^4} \, \frac{1}{\tilde Q^2}
 \; \biggr\} 
 \; 
 \approx \;
 1 - \frac{0.59777}{\beta}
 \;, 
\ea 
with $\tilde q^2, \tilde Q^2$ defined in \eq\nr{qQ_defs}.  
The renormalization factor is 
then $c_2^2 \approx 0.83 ... 0.85$, i.e.\ quite modest. The results obtained 
after this correction ($\equiv G_\rmi{\,imp,Z}$)
are shown in \fig\ref{fig:physical}(left). Of course, systematic
(non-perturbative) renormalization would be highly desirable~\cite{rs_review}.

%
\section{Physical interpretation}

The correlator $G_\rmi{\,E}$ was computed at 
next-to-leading order (NLO) in continuum 
in ref.~\cite{rhoE}. A comparison with our lattice results, 
improved and renormalized as explained above, is shown 
in \fig\ref{fig:physical}(left). We observe a clear 
enhancement over the NLO prediction over a wide 
Euclidean time interval. 
Subtracting the two results, one could in principle attempt 
a model-independent analytic continuation of the non-perturbative
surplus, for instance along the lines of ref.~\cite{analytic}. 
Unfortunately, at the current stage, our 
resolution is not sufficient ($\sim 10^{-3}$) for this task. 
Therefore we resort to a rough model-dependent interpretation
of the data in the following. 

%
\begin{figure}[t]

\centerline{%
 \epsfysize=7.0cm\epsfbox{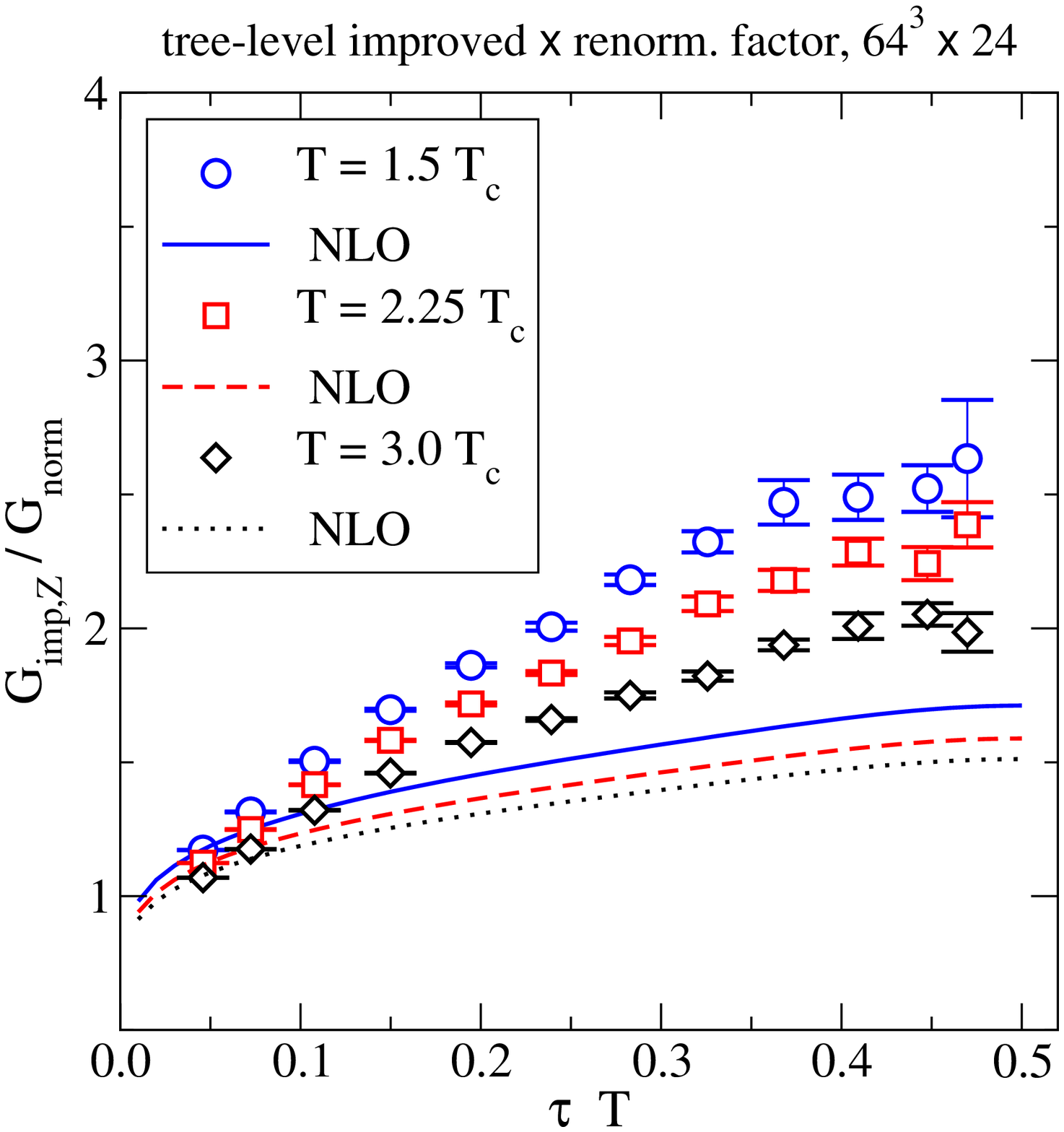} 
 \hspace*{0.5cm}%
 \epsfysize=7.0cm\epsfbox{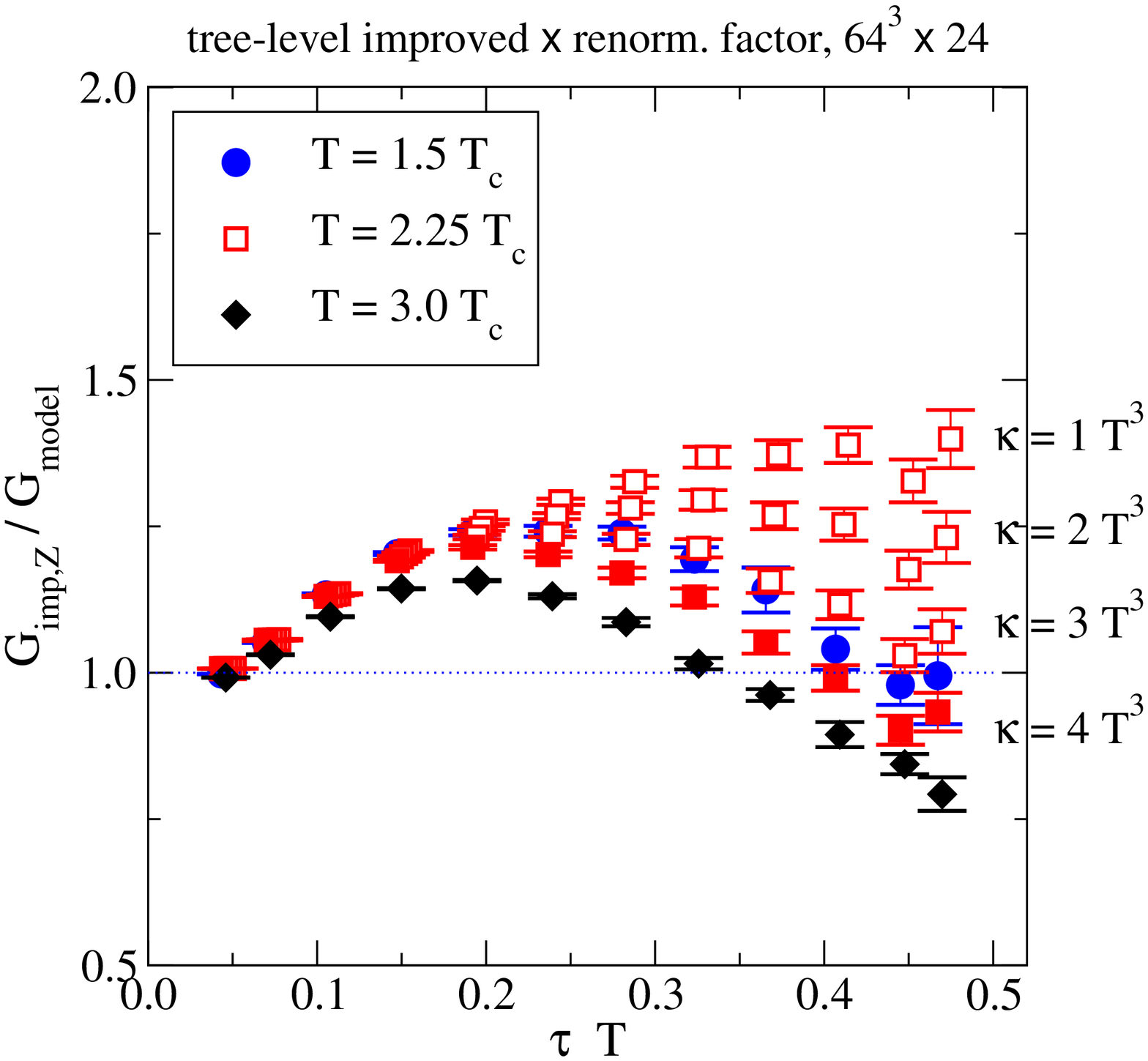} 
}

\caption[a]{\small 
 Left: Comparison of lattice with the NLO weak-coupling 
 expansion~\cite{rhoE}. 
 Right: Like on the left but normalized to the model of 
 \eq\nr{model}. Closed symbols correspond to $\kappa = 4 T^3$, 
 open ones to $T = 2.25\Tc$.
} 
 \la{fig:physical}
\end{figure}
%

Most transport coefficients are related to quantities which are 
conserved in the free limit, and therefore arise in connection
with a narrow transport peak. This yields a contribution to 
the Euclidean correlator which is almost constant in $\tau$. 
Since the full Euclidean correlator diverges at  
short distances and is essentially perturbative there, 
ratios such as $G_\rmi{\,lat}/G_\rmi{\,NLO}$
should only be enhanced around the middle of the 
Euclidean time interval (cf.\ e.g.\ refs.~\cite{ding2,ding1}).
This is clearly {\em not} the case in \fig\ref{fig:physical}(left), 
and a narrow transport peak can be excluded, as expected~\cite{eucl,rhoE}. 

For a potentially more realistic model, we take inspiration from 
non-perturbative computations of the colour-electric correlator in 
other theories, one of them strongly coupled ${\mathcal N}=4$
Super-Yang-Mills in the large-$\Nc$ limit~\cite{ct,Gubser:2006nz}, 
another classical lattice gauge theory with $\Nc = 3$~\cite{mink}. 
Both suggest $\rho_\rmi{\,E}(\omega)/\omega$ to essentially 
flatten off below some frequency scale. Therefore, we define
\be
 \rho_\rmi{\,model}(\omega) 
 \equiv \mathop{\mbox{max}}
 \Bigl\{ \rho_\rmi{\,NLO}(\omega) , \frac{\omega \kappa}{2 T} \Bigr\} 
 \;, \la{model}
\ee
with the free parameter $\kappa$ representing directly the 
momentum diffusion coefficient according to \eq\nr{intercept}, 
and compute the corresponding Euclidean correlator from 
\be
 G_\rmi{\,model}(\tau) \equiv 
 \int_0^\infty
 \frac{{\rm d}\omega}{\pi} \rho_\rmi{\,model}(\omega)
 \frac{\cosh \left(\frac{1}{2} - \tau T \right)\frac{\omega}{T} }
 {\sinh\frac{\omega}{2 T}} 
 \;. \la{int_rel} 
\ee 
Results are shown in \fig\ref{fig:physical}(right).
We observe that despite extending its influence to larger frequencies
than a narrow transport peak, the model does not capture 
the full shape of the Euclidean correlator; probably, extra 
``power'' needs to be added to $\rho(\omega)$ at even larger 
frequencies.

Nevertheless, aiming at a match around the middle of the Euclidean
time interval, we can read from \fig\ref{fig:physical}(right) that 
 $\kappa \sim 4 T^3$ at $T = 1.5\Tc$; 
 $\kappa \sim 3.5 T^3$ at $T = 2.25\Tc$; and 
 $\kappa \sim 2.5 T^3$ at $T = 3.0\Tc$. 
The NLO weak-coupling expansion yields values 
$\kappa \sim 2 T^3$~\cite{chm2} in this temperature range, 
and classical lattice gauge theory suggests that this 
could be an underestimate~\cite{mink}. 
Converting $\kappa$ to the ``usual'' 
diffusion coefficient $D = 2 T^2/\kappa$, we obtain 
$D \sim (0.5 ... 0.8)/T$, which might lie 
in a phenomenologically acceptable range (cf.\ e.g.~refs.~\cite{ct,vH}). 
In ref.~\cite{ding2}, values $D \sim 0.3/T$
were cited for the charm case. It is interesting that, 
despite the very rough nature of all of these estimates, 
a somewhat consistent picture appears to emerge. 
%
%

%
\section{Conclusions}

As the results in fig.~\ref{fig:physical} show, 
it is possible to obtain results for the colour-electric 
correlator which unambiguously demand
an enhancement of the non-perturbative interactions over the 
weak-coupling description and which furthermore strongly constrain
the shape of the corresponding spectral function. Of course, 
for quantitative statements, careful infinite-volume and 
continuum extrapolations need to be taken, 
statistical uncertainties should be further reduced, 
and a more careful modelling of the spectral shape is called for. 

%
\section*{Acknowledgements}

 We thank H.~Meyer for providing us 
 with lattice data from ref.~\cite{hbm} for comparison.
 A.F.\ was partly supported by 
 the DFG International Graduate School 
 {\em Quantum Fields and Strongly Interacting Matter}, 
 and M.L.\ and J.L.\ by
 the BMBF under the project
 {\em Heavy Quarks as a Bridge between
      Heavy Ion Collisions and QCD}.

%

\end{document}